\documentclass[superscriptaddress,amsmath,amssymb,aps,reprint,prb,twocolumn]{revtex4-2}

\usepackage{graphicx}% Include figure files
\usepackage{dcolumn}% Align table columns on decimal point
\usepackage{bm}% bold math
\usepackage{hyperref}% add hypertext capabilities
\usepackage[normalem]{ulem}
\hypersetup{
	citecolor = blue,
	colorlinks = true,
	urlcolor = blue
}
\usepackage{xcolor}
\usepackage{braket} 

\begin{document}

\title{\textbf{Electronic driving of torsional modes in carbon nanotubes via orbital-phonon coupling}}

\author{Koji Yamada}
\affiliation{Institute for Solid State Physics, University of Tokyo, Kashiwa 277-8581, Japan}

\author{Wataru Izumida}
\affiliation{Department of Physics, Tohoku University, Sendai 980-8578, Japan}

\author{Mamoru Matsuo}
\affiliation{Kavli Institute for Theoretical Sciences, University of Chinese Academy of Sciences, Beijing, 100190, China.}
\affiliation{CAS Center for Excellence in Topological Quantum Computation, University of Chinese Academy of Sciences, Beijing 100190, China}
\affiliation{Advanced Science Research Center, Japan Atomic Energy Agency, Tokai, 319-1195, Japan}
\affiliation{RIKEN Center for Emergent Matter Science (CEMS), Wako, Saitama 351-0198, Japan}

\author{Takeo Kato}
\affiliation{Institute for Solid State Physics, University of Tokyo, Kashiwa 277-8581, Japan}

\date{\today}

\begin{abstract}
We theoretically investigate electronic driving of a torsional mode in a suspended carbon nanotube quantum dot. 
The valley states in a nanotube carry opposite orbital magnetic moments along the tube axis, and intervalley mixing forms valley-mixed eigenstates whose splitting can be controlled by an axial magnetic field.
We formulate an orbital-rotation coupling between the torsional angular velocity and the valley orbital angular momentum. 
Using a master equation for sequential tunneling through normal-metal electrodes, we calculate the steady-state current and the phonon distribution of the torsional mode. 
When the valley-state splitting matches the torsional phonon energy, orbital transitions can drive the mode out of equilibrium. The contacts control both phonon pumping and its electrical readout, allowing substantial excitation with a positive, vanishing, or negative current response. The mechanism requires no intrinsic spin or valley polarization of the electrodes, but uses state-dependent tunneling at the contacts.
\end{abstract}

\maketitle

\section{Introduction}
\label{sec:introduction}

Carbon nanotubes (CNTs) are useful platforms for nanoelectromechanical systems because their small mass, large stiffness, and high aspect ratio allow mechanical resonances to be combined with electronic transport at the nanoscale~\cite{Craighead2000,Sapmaz2003,Sazonova2004,PootZant2012}. 
In suspended CNT devices, the nanotube can also work as a quantum dot, and discrete electronic states and mechanical motion can be controlled in the same structure~\cite{Huttel2009,Steele2009,Lassagne2009,Laird2015}. 
Such devices allow one to study electromechanical coupling between internal electronic degrees of freedom and mechanical motion.

In most CNT electromechanical devices, electrical control of mechanical motion is based on capacitive forces or deformation-potential coupling, both of which couple naturally to displacement-like vibrational modes~\cite{Ekinci2005,Steele2009,Lassagne2009}. 
Torsional motion is qualitatively different because its generalized coordinate is an angular displacement around the tube axis, and the corresponding drive requires a torque rather than a transverse force. 
Although torsional motion of CNT devices has been demonstrated using torsion springs and torsional resonators~\cite{Papadakis2004,Hall2008,Hall2006}, efficient coupling between electronic transport and a torsional mode is still not straightforward. 
In particular, conventional electron-vibron coupling to torsional modes in suspended CNT quantum dots is expected to be weak compared with the coupling to stretching or bending modes~\cite{Mariani2009}. 
We therefore consider a torque drive based on angular-momentum transfer rather than force-displacement coupling.

Carbon nanotube quantum dots provide a natural electronic degree of freedom for such a mechanism. 
The two valley states correspond to clockwise and counterclockwise motion around the tube circumference and carry opposite orbital magnetic moments along the tube axis~\cite{Minot2004,Jarillo-Herrero2005,Laird2015}. 
These orbital magnetic moments can be much larger than the spin magnetic moment, and an axial magnetic field shifts the two valley levels in opposite directions. 
In realistic quantum dots, disorder and tube-end scattering mix the two valley states and produce an avoided crossing near zero axial magnetic field~\cite{Jarillo-Herrero2005,Kuemmeth2008,Laird2015}. 
The valley-mixed eigenstates retain the orbital-angular-momentum character of the original valley states and allow transitions between the two eigenstates.

Our approach is motivated by our recent study of torsional-mode excitation driven by spin angular momentum in a CNT quantum dot~\cite{Yamada2026}. 
In that system, spin-rotation coupling enables spin-flip tunneling cycles to create torsional phonons, and a resonant current signal appears when the spin splitting matches the torsional-mode energy. 
Here we extend this angular-momentum-transfer mechanism from spin to the valley orbital degree of freedom. 
Because the orbital angular momentum associated with the valley states can be much larger than the spin angular momentum, orbital-rotation coupling can provide a stronger channel for driving torsional motion. 
The orbital mechanism also works with normal-metal electrodes rather than ferromagnetic spin-selective contacts. 
Normal electrodes generate a finite background current because they are not valley-selective, but they simplify the setup; the torsional-mode excitation is detected as a resonant modulation of the current on this background.
The normal-metal contacts play two roles: they control the population imbalance that supplies phonon pumping, and they determine how the internal orbital transitions appear in the current. Consequently, substantial torsional excitation can coexist with a positive, vanishing, or negative electrical response.
In this paper, we propose a mechanism for driving a quantized torsional mode of a suspended CNT by the orbital angular momentum of electrons in a CNT quantum dot. 
We formulate the orbital-rotation coupling between the valley-mixed dot states and the torsional angular velocity, and then derive transition rates for sequential tunneling and phonon-assisted orbital transitions.
Using a master equation for the coupled dot--phonon dynamics, we calculate the steady-state current and phonon distribution under a finite bias voltage. 
We show that the current exhibits a contact-dependent resonant response when the energy splitting between the valley-mixed states matches the torsional phonon energy.
We also show that this resonance is accompanied by nonequilibrium excitation of the torsional mode.

The remainder of the paper is organized as follows. In Sec.~\ref{sec:model}, we introduce the model Hamiltonian of a CNT quantum dot, a clamped CNT torsional mode, and an orbital-rotation coupling.
In Sec.~\ref{sec:formulation}, we present the transition rates and the master equation. 
In Sec.~\ref{sec:estimate}, we evaluate experimental feasibility by estimating model parameters.
In Sec.~\ref{sec:results}, we discuss the resonant behavior of the steady-state current and the phonon distribution. 
In Sec.~\ref{sec:discussion}, we briefly present the relation to previous works on spin-rotation coupling.
Finally, we summarize the results in Sec.~\ref{sec:summary}.

\section{Model}
\label{sec:model}

\begin{figure}[tb]
\centering
\includegraphics[width=8.5cm]{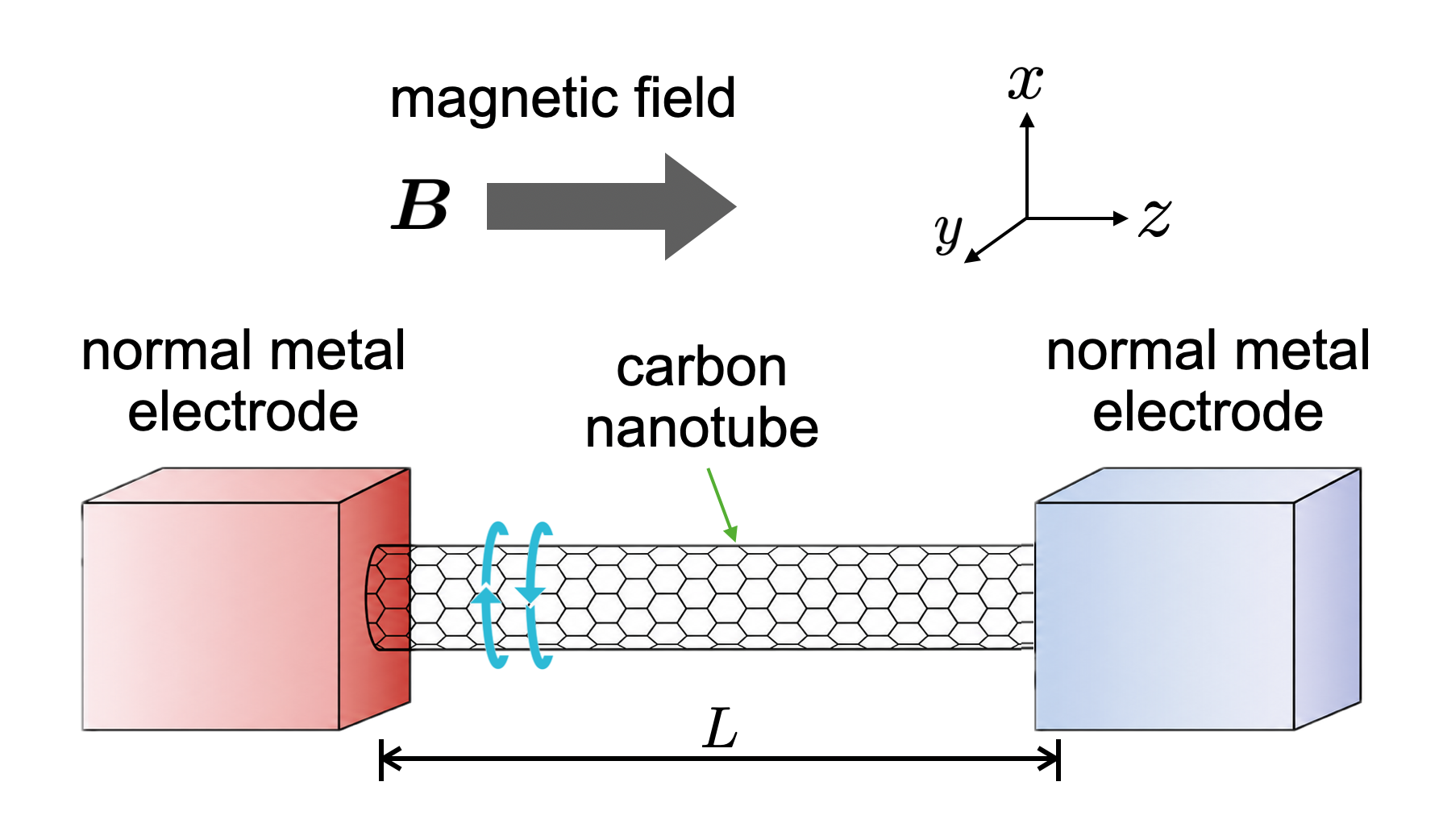}
\caption{Schematic of the model: A suspended CNT of length \( L \) is connected to normal-metal electrodes.
}
\label{fig:model}
\end{figure}

We consider a suspended single-wall CNT of length \(L\), whose ends are rigidly clamped by normal-metal electrodes, as illustrated in Fig.~\ref{fig:model}. 
We set the laboratory coordinates \(xyz\) so that the $z$ axis is along the CNT axis.
A static magnetic field \({\bm B}=B\hat z\) is applied along the CNT axis.
The system Hamiltonian is given by
\begin{align}
H
&= H_{\mathrm{dot}} + H_{\mathrm{lead}} + H_{\mathrm{lead-dot}}
\notag\\
&\quad + H_{\mathrm{ph}} + H_{\mathrm{env}} + H_{\mathrm{ph-env}}
+ H_{\mathrm{OR}} ,
\label{eq:H_total}
\end{align}
where \(H_{\mathrm{dot}}\), \(H_{\mathrm{lead}}\), and \(H_{\mathrm{lead-dot}}\) describe the CNT quantum dot,
the electrodes, and the electron tunneling between the electrodes and the dot, respectively. The torsional phonon mode is described by \(H_{\mathrm{ph}}\), whereas \(H_{\mathrm{env}}\) and \(H_{\mathrm{ph-env}}\) model its coupling to the surrounding environment.
The last term $H_{\mathrm{OR}}$ describes the orbital-rotation coupling between the torsional mode and the electron in the CNT quantum dot.
In the following subsections, we present explicit forms of each Hamiltonian.

\subsection{CNT quantum dot}

The Hamiltonian of the CNT quantum dot is given by
\begin{align}
H_{\mathrm{dot}} = H_0 + H_{\mathrm{int}},
\end{align}
where \(H_0\) is the single-particle Hamiltonian without electron-phonon coupling, and \(H_{\mathrm{int}}\) represents the Coulomb interaction.

Let us first consider the single-particle Hamiltonian \(H_0\).
For the CNT quantum dot, we consider four  
states denoted by \( |\tau s \rangle\),  where \( \tau = {\rm K}, {\rm K}'\)
denotes the valley degree of freedom and 
$s=\uparrow,\downarrow$
describes the electron spin.
We parameterize the intervalley-mixing term by the anticrossing gap  \(\Delta_{{\rm KK}'}>0\) and a phase \(\chi\).

The single-electron Hamiltonian is then written as
\begin{align}
H_0
=&-\frac{g\mu_{\rm B} B}{2}\hat{s}_z
-\mu_{\rm orb}B\hat{\tau}_z
-\frac{\Delta_{\rm SO}}{2}\hat{s}_z\hat{\tau}_z
\nonumber\\
&-\frac{1}{2}
\left(
\Delta_{{\rm KK}'} e^{-i\chi} \hat{\tau}_{+}
+ \Delta_{{\rm KK}'} e^{i\chi} \hat{\tau}_{-}
\right),
\label{eq:H_v_sigma_full}
\end{align}
where $g$ is the electron $g$-factor, \(\mu_{\rm B}\) is the Bohr magneton, \(\mu_{\rm orb}\) is the orbital magnetic moment due to the electron's circular motion around the CNT axis, 
$B$ is the external magnetic field applied along the tube axis,
$\Delta_{\rm SO}$ is the strength of the spin-orbit coupling, $\hat{\tau}_a$ and $\hat{s}_a$ ($a=x,y,z$) are the Pauli matrices for the valley and spin degrees of freedom, and $\hat{\tau}_{\pm}=(\hat{\tau}_x\pm i\hat{\tau}_y)/2$.
The overall sign of the intervalley-mixing term is a matter of convention and can be changed by redefining the phase of one of the valley basis
states.

\begin{figure}[t]
\centering
\includegraphics[width=8cm]{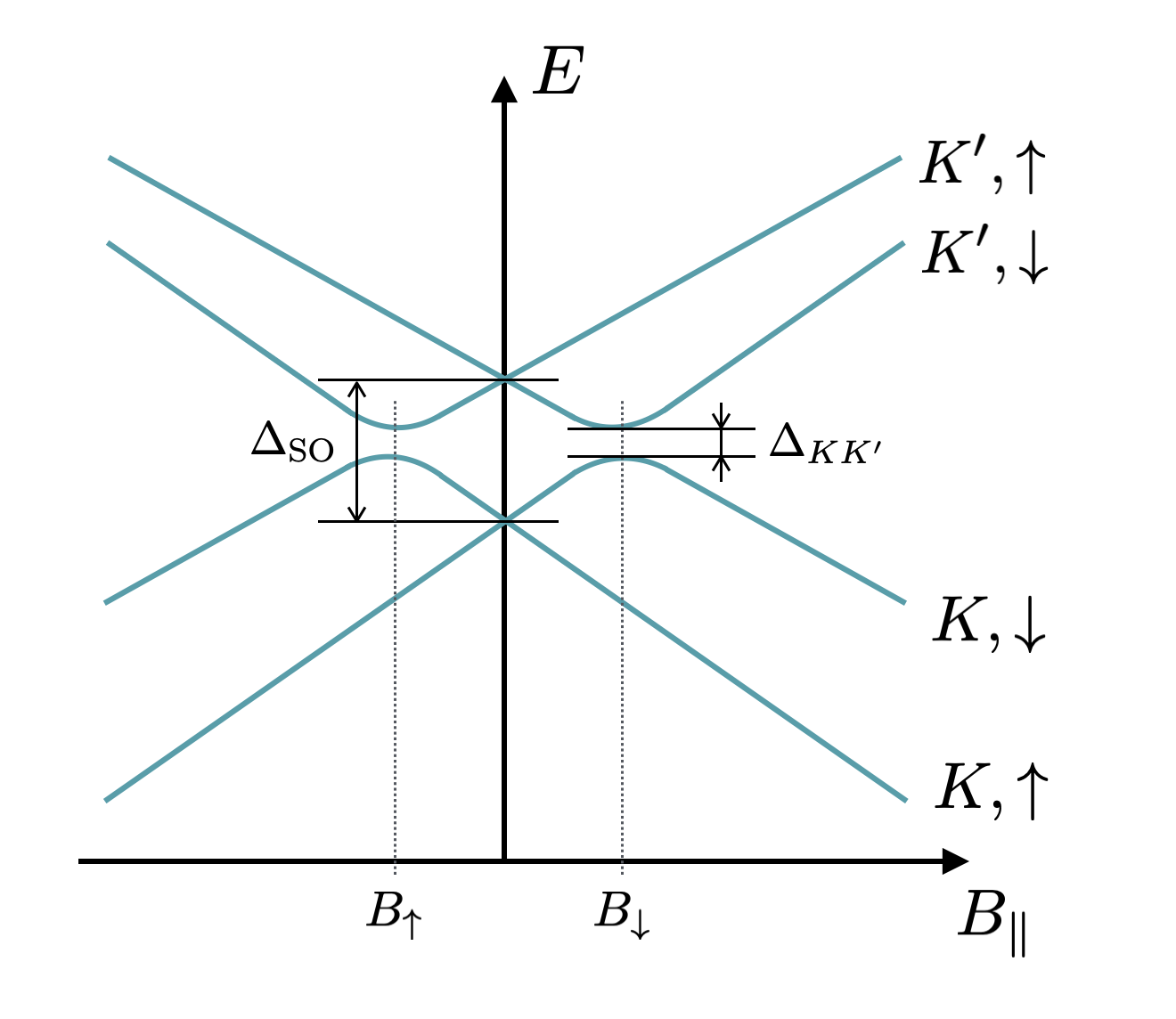}
\caption{Schematic of the single-electron energy levels as a function of the external magnetic field $B$. $B_s$ ($s = \uparrow, \downarrow$) denotes the magnetic field at which the anticrossing of the energy levels with the same spin occurs.}
\label{fig:levels}
\end{figure}

The single-electron energy levels are schematically illustrated in Fig.~\ref{fig:levels}.
We note that this energy-level structure has been observed experimentally~\cite{Kuemmeth2008}.
At zero magnetic field, the spin-orbit interaction lifts the fourfold degeneracy of the states
$|\tau s\rangle$, splitting them into two Kramers doublets, each consisting
of a pair of time-reversal partners,
$(K,\uparrow)$ and $(K',\downarrow)$ or
$(K,\downarrow)$ and $(K',\uparrow)$.
The magnetic field shifts these energy levels through the Zeeman energy for the spin and orbital angular momenta.
Since \( \mu_{\rm orb} \) is larger than \( g\mu_{\rm B} \), the slope of the energy diagram in Fig.~\ref{fig:levels} is mainly determined by the orbital degree of freedom $\tau= {\rm K}, {\rm K}'$, 
while the spin degree of freedom changes the slope slightly.
Near the crossing point between the states 
$|{\rm K} s \rangle$ and $|{\rm K}' s \rangle$, 
the valley mixing term opens an energy gap of $\Delta_{{\rm KK}'}$ at $B=B_{s}$ where 
\begin{align}
B_{s} = - s \frac{\Delta_{\rm SO}}{2\mu_{\rm orb}},
\end{align}
where $s=+1$ ($-1$) for spin $\uparrow$ ($\downarrow$).

In our work, we focus on this crossing point and consider only the two states $|{\rm K} s \rangle$
and $|{\rm K}' s \rangle$ for a fixed $s$,
assuming that the energies of the other two states lie far from the chemical potentials in the electrodes.
To simplify the discussion, we focus on the two states with spin $\downarrow$ and suppress the spin index $s$ hereafter.
Introducing the detuning $\delta B=B-B_{\downarrow}$ and omitting the spin Zeeman term, we obtain the effective two-state Hamiltonian in the valley basis as
\begin{align}
H_0 =
\begin{pmatrix}
-\mu_{\rm orb}\delta B &
-\dfrac{\Delta_{{\rm KK}'}}{2}e^{-i\chi}\\[1mm]
-\dfrac{\Delta_{{\rm KK}'}}{2}e^{i\chi} &
\mu_{\rm orb}\delta B
\end{pmatrix}.
\label{eq:H_v_eff}
\end{align}
The two eigenenergies are given by
\begin{align}
E_{\pm} = \pm \sqrt{(\mu_{\rm orb}\delta B)^2+\Delta_{{\rm KK}'}^2/4},
\end{align} 
giving the energy splitting $\Delta E=E_+-E_-$.
We define the mixing angle $\phi$ ($0\leq\phi\leq\pi$) by
\begin{align}
\cos\phi &= \frac{\mu_{\rm orb}\delta B}{E_+}, \\
\sin\phi &= \frac{\Delta_{{\rm KK}'}}{2E_+} .
\label{DefMixingAngle}
\end{align}
The corresponding eigenstates are given by
\begin{align}
|-\rangle &=
\cos(\phi/2)|{\rm K}\rangle
+e^{i\chi}\sin(\phi/2)|{\rm K}'\rangle,
\label{eq:EigenMinus}
\\
|+\rangle &=
-\sin(\phi/2)|{\rm K}\rangle
+e^{i\chi}\cos(\phi/2)|{\rm K}'\rangle.
\label{eq:EigenPlus}
\end{align}

Next, we consider the effect of the Coulomb interaction.
As already stated, we consider transport through an active pair of spin-down valley states near $B=B_{\downarrow}$ and neglect the other states including the unoccupied spin-up level $(K',\uparrow)$ and the occupied spin-up level $(K,\uparrow)$, by assuming the energies necessary for adding (removing) one electron to these unoccupied (occupied) states to be much larger than the source-drain bias.
Furthermore, we assume strong Coulomb interaction, for which the charging energy also prevents simultaneous occupation of both active  states by additional electrons.
Therefore, we retain only the reference charge state $\ket{0}$ and the two states $\ket{\nu}$ (\(\nu=\pm\)) obtained by adding one electron to the active levels.
We note that $\ket{0}$ denotes the absence of an additional electron in the active quantum levels, rather than an empty dot.
Within the constant-interaction approximation, the charging energy required to add an electron is the same for the two active levels. We include this common energy in the addition-energy offset and omit $H_{\mathrm{int}}$ from the projected Hamiltonian. We still restrict the active doublet to at most one additional electron.
The energies entering the tunneling rates discussed in Sec.~\ref{sec:TransitionRate} are the addition energies relative to the reference charge state. We choose the gate-controlled addition-energy offset to vanish, so that these energies coincide with $E_\pm$.

\subsection{Electrodes}

\begin{figure}[t]
\centering
\includegraphics[width=8.5cm]{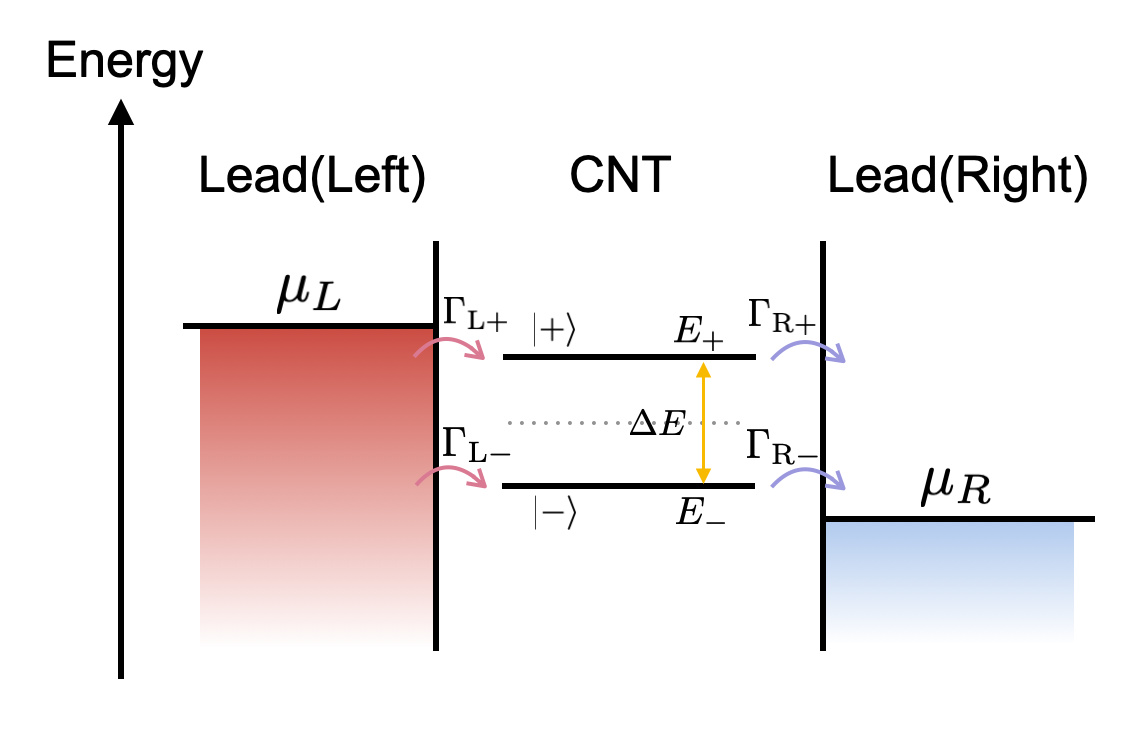}
\caption{Energy-level diagram of the CNT quantum dot. \label{fig:energy_diagram}}
\end{figure}

The two electrodes are modeled as noninteracting electron systems, whose Hamiltonian is given by 
\begin{align}
H_{\mathrm{lead}} &= \sum_{r={\rm L},{\rm R}} \sum_{{\bm k}} (\epsilon_{\bm k}-\mu_r) c_{r{\bm k}}^\dagger c_{r{\bm k}}, 
\end{align}
where $c_{r {\bm k}}$ denotes an annihilation operator for electrons with wave vector ${\bm k}$ in the electrode $r = {\rm L},{\rm R}$, $\epsilon_{\bm k}$ is the electron energy, and $\mu_r$ is the chemical potential of the electrode $r$.
We assume that the two energy levels of the CNT are located between the chemical potentials of the two electrodes, as shown in Fig.~\ref{fig:energy_diagram}.
We have suppressed the spin index since the spin state that contributes to transport through the CNT is fixed.

The coupling between the CNT quantum dot and the electrodes is described by the tunnel Hamiltonian
\begin{align}
H_{\mathrm{lead-dot}} &= 
\sum_{r={\rm L},{\rm R}} \sum_{\bm k} \sum_{\tau 
= {\rm K},{\rm K}'} V_{r{\bm k}\tau} d_{\tau}^\dagger c_{r {\bm k}} 
+ {\rm H.c.},
\end{align}
where $d_\tau$ denotes an annihilation operator for an electron with the valley index $\tau$ and $V_{r{\bm k}\tau}$ denotes a tunneling amplitude. Since the valley-mixed dot eigenstates are coherent superpositions of
the ${\rm K}$ and ${\rm K}'$ states, the relative amplitudes and phases
of $V_{r{\bm k}{\rm K}}$ and $V_{r{\bm k}{\rm K}'}$ can affect
the tunneling rates to the two dot eigenstates.
We retain this possibility and derive the corresponding tunneling rates
in Sec.~\ref{sec:formulation}.

\subsection{Torsional mode}
\label{sec:TorsionalMode}

The torsional mode is described by the Hamiltonian~\cite{Suzuura2002,Yamada2026}
\begin{align}
H_{\rm ph} = \hbar \omega_0 a^\dagger a ,
\end{align}
where \( a \) and \( a^\dagger \) are bosonic annihilation and creation operators. The natural frequency is given by
\begin{align}
\omega_0 =
\frac{\pi}{L}\sqrt{\frac{GS}{\rho}} ,
\end{align}
where \( L \) is the CNT length, \(G\) is the shear modulus, \(S=2\pi Rd\) is the effective cross-sectional area of the CNT wall, \( R \) is the CNT radius, \(d\) is the wall thickness, and \(\rho\) is the mass per unit length.

To incorporate the damping of the torsional mode, we introduce a phonon bath coupled to the torsional oscillator, whose Hamiltonian is given by
\begin{align}
H_\mathrm{env} 
&= \sum_{\bm q}\hbar \omega_{\bm q}
\left(b_{\bm q}^\dagger b_{\bm q}+\frac{1}{2}
\right),\\
H_\mathrm{ph-env}
&= \sum_{\bm q}\kappa\left(a^\dagger b_{\bm q}+ a b^\dagger_{\bm q}
\right).
\end{align}
Here, \(b_{\bm q}\) and \(b_{\bm q}^\dagger\) are the annihilation and creation operators for phonons with wave vector \(\bm q\), respectively, and \(\kappa\) characterizes the coupling between the CNT torsional mode and the phonon bath.

\subsection{Orbital-rotation coupling}

We use the orbital part of the gyromagnetic coupling to drive the torsional mode.
The relevant interaction is the orbital-rotation coupling between an electron in the CNT and the torsional mode.
We briefly summarize the derivation of this coupling Hamiltonian.

The fundamental torsional mode is described by the twist-angle operator
\begin{align}
\hat{\theta}(z,t)
&=
\sqrt{\frac{2}{L}}\sin\frac{\pi z}{L}\,\hat{\theta}(t), \\
\hat{\theta}(t)
&=
\sqrt{\frac{\hbar}{2\rho R^2\omega_0}}
\left(a e^{-i\omega_0 t} + a^\dagger e^{i\omega_0 t}\right),
\end{align}
where \(a\) and \(a^\dagger\) are the annihilation and creation operators for the torsional phonon introduced in Sec.~\ref{sec:TorsionalMode}.
This torsional motion modulates the positions of the impurities in the CNT and makes the intervalley-scattering Hamiltonian time-dependent.
For a uniformly rotating potential, this time dependence can be removed by transforming to a co-rotating frame.
The electron Hamiltonian then acquires an inertial term coupling the angular velocity to the total angular momentum~\cite{Hehl1990,Matsuo2011,Ominato2022},
\begin{align}
H_{\rm dot} \rightarrow H_{\rm dot} - \bm{\omega}\cdot(\bm L + \bm S),
\end{align}
where $\bm L$ and $\bm S$ are the orbital and spin angular momenta.
The spin part, \(-\bm{\omega}\cdot\bm S\), gives the
spin-rotation coupling, which has been used to excite the torsional mode~\cite{Yamada2026}. 
Here we focus on the orbital part and describe the dot--phonon coupling by the Hamiltonian
\begin{align}
H_{\rm OR} = -\bm{\omega}\cdot\bm L .
\label{eq:OR_basic}
\end{align}
Let $P=|K\rangle\langle K|+|K'\rangle\langle K'|$
project onto the valley subspace of the active dot orbital in the selected spin sector. We approximate the two valley states as carrying equal and opposite axial orbital angular momenta~\cite{Ajiki1993,Minot2004,Laird2015} and neglect the intervalley matrix element of $\hat L_z$~\cite{Izumida2016,Laird2015}. The projected operator then reads
\begin{align}
P\hat L_zP \simeq \hbar l\,\hat\tau_z,
\end{align}
where $l=\langle K|\hat L_z|K\rangle/\hbar$ is a dimensionless effective parameter.

For a torsional vibration, the local lattice rotation is spatially nonuniform.
We approximate the coupling by a local extension of the axial rotation term, assuming that the relevant scatterers follow the local lattice motion. Additional strain-induced couplings and changes in the contact tunneling amplitudes are neglected.
The orbital-rotation coupling is written as
\begin{align}
H_{\rm OR} = -\int d^3\bm r\, \bm{\omega}(\bm r)\cdot\bm{\mathcal L}(\bm r),
\label{eq:OR_basic2}
\end{align}
where $\bm{\mathcal L}(\bm r)$ is the electronic orbital angular-momentum density, satisfying $\int d^3\bm r\,\bm{\mathcal L}(\bm r)=\bm L$~\footnote{Coupling between lattice vorticity and the valley degree of freedom has been studied in two-dimensional Dirac materials driven by classical surface acoustic waves (Ref.~\cite{Ominato2022}). Here we consider the reverse process, in which electron transport through a quantum dot pumps a quantized one-dimensional torsional phonon mode.}.
The local angular velocity is related to the vorticity of the lattice velocity field ${\bm v}({\bm r})$ as
\begin{align}
\bm{\omega}(\bm r)
= \frac{1}{2}\nabla\times\bm v(\bm r).
\end{align}
Projecting Eq.~\eqref{eq:OR_basic2} onto the active valley subspace, we assume that the two valley states share a uniform axial probability density $1/L$ over the vibrating segment of a CNT of length $L$. 
The coupling then involves the spatial average of the local axial angular velocity.
Retaining this component and substituting the fundamental torsional mode, we obtain
\begin{align}
H_{\rm OR} = i g_{\rm OR}(a-a^\dagger)\hat{\tau}_z,
\label{eq:H_OR_final}
\end{align}
where the coupling constant is given by
\begin{align}
g_{\rm OR} = \frac{2\hbar l}{\pi R} \sqrt{\frac{\hbar\omega_0}{L\rho}}.
\label{eq:g_OR}
\end{align}
Within the projected orbital model specified above, the remaining derivation follows the procedure used for the spin-rotation coupling in Ref.~\cite{Yamada2026}.

\section{Formulation}
\label{sec:formulation}

We calculate electron transport through the CNT using a master-equation approach.
We work in the sequential-tunneling regime and treat $H_{\rm lead-dot}$, $H_{\rm ph-env}$, and $H_{\rm OR}$ perturbatively within the Born--Markov approximation.
The relevant states are denoted by $(\alpha,m)$, where $\alpha=0,+,-$ labels the electronic state of the dot and $m=0,1,2,\ldots$ is the phonon number of the torsional mode.
In the following, we first derive the transition rates associated with electron tunneling, mechanical damping, and the orbital-rotation coupling, and then formulate the master equation for the occupation probabilities $P_{\alpha,m}$.

\subsection{Transition rates}
\label{sec:TransitionRate}

Before giving the explicit rates, we summarize the transport cycle described by the master equation. 
Electron transport proceeds through sequential tunneling between the normal electrodes and the valley-mixed dot eigenstates (see Fig.~\ref{fig:energy_diagram}). 
The electron tunneling between the electrodes and the CNT quantum dot does not change the torsional phonon number. The phonon number changes through the orbital-rotation coupling, which induces transitions between the two dot eigenstates while creating or annihilating a torsional phonon.
Near the resonance condition \(E_{+}-E_{-}\simeq \hbar\omega_0\), sequential tunneling and orbital-rotation-induced transitions form a cycle that generates nonequilibrium phonons in the torsional mode.

We first consider electron tunneling between the electrodes and the CNT quantum dot.
The tunneling Hamiltonian $H_{\rm lead-dot}$ couples the electrode states to the ${\rm K}$ and ${\rm K}'$ components of the dot states. 
To describe this interference systematically, we introduce a
tunneling-rate matrix in the valley subspace.
Within the wide-band approximation, its matrix elements are defined as
\begin{align}
\bm{\Gamma}_r &= \begin{pmatrix}
\Gamma_r^{{\rm KK}} & \Gamma_r^{{\rm KK'}} \\
\Gamma_r^{{\rm K'K}} & \Gamma_r^{{\rm K'K'}} \\
\end{pmatrix} , \quad (r ={\rm L}, {\rm R}), \\
\Gamma_r^{\tau\tau'}
&=
\frac{2\pi}{\hbar}
\sum_{\bm k}
V_{r{\bm k}\tau}
V^{*}_{r{\bm k}\tau'}
\delta(\epsilon_{\bm k}-\epsilon_{\rm F}).
\label{eq:Gamma_matrix_definition}
\end{align}
The diagonal elements describe the tunneling strengths to the individual valley components, whereas the off-diagonal elements describe the interference between the tunneling amplitudes to the ${\rm K}$ and ${\rm K}'$ components.
Since $\bm{\Gamma}_r$ is positive semidefinite, we can parameterize it as
\begin{align}
\bm{\Gamma}_r
=
\begin{pmatrix}
\Gamma_{r{\rm K}}
&
p_r\sqrt{\Gamma_{r{\rm K}}\Gamma_{r{\rm K}'}}
e^{-i\vartheta_r}
\\
p_r\sqrt{\Gamma_{r{\rm K}}\Gamma_{r{\rm K}'}}
e^{i\vartheta_r}
&
\Gamma_{r{\rm K}'}
\end{pmatrix},
\label{eq:Gamma_matrix}
\end{align}
where $\Gamma_{r\tau} = \Gamma_{r}^{\tau\tau}$ ($\tau={\rm K},{\rm K}'$) are the diagonal elements, $p_r$ ($0\leq p_r\leq1$) characterizes the degree of valley coherence retained by the contact, and $\vartheta_r$ denotes the relative phase between the tunneling amplitudes to the two valley components.
The limit $p_r=0$ corresponds to completely incoherent tunneling, whereas $p_r=1$ describes a fully coherent single tunneling channel. We emphasize that $p_r\neq0$ does not imply that the metallic electrode itself is valley polarized.
The electrodes are ordinary normal metals and do not possess a valley degree of freedom analogous to that of the CNT. Rather, valley coherence arises at the contact because the same electrode state can tunnel coherently into both valley components of the CNT wave function. Consequently, a spatially local contact can preferentially couple to a particular superposition of the CNT valley states~\cite{GroveRasmussen2012,Donarini2019}.

Using the eigenstates given in Eqs.~\eqref{eq:EigenMinus} and \eqref{eq:EigenPlus}, the projected rates are
\begin{align}
\Gamma_{r-}
=&\,
\Gamma_{r{\rm K}}\cos^2(\phi/2)
+\Gamma_{r{\rm K}'}\sin^2(\phi/2)
\nonumber\\
&+
p_r\sqrt{\Gamma_{r{\rm K}}\Gamma_{r{\rm K}'}}
\sin\phi\cos\Phi_r ,
\label{Gammaminus}
\\
\Gamma_{r+}
=&\,
\Gamma_{r{\rm K}}\sin^2(\phi/2)
+\Gamma_{r{\rm K}'}\cos^2(\phi/2)
\nonumber\\
&-
p_r\sqrt{\Gamma_{r{\rm K}}\Gamma_{r{\rm K}'}}
\sin\phi\cos\Phi_r ,
\label{Gammaplus}
\end{align}
where
\begin{align}
\Phi_r=\chi-\vartheta_r
\end{align}
is the gauge-invariant relative phase between the intervalley mixing in the dot and the tunneling amplitude at contact $r$. Only this relative phase enters the observable tunneling rates; the individual phases $\chi$ and $\vartheta_r$ depend on the phase convention chosen for the valley basis.
To isolate the effect of valley-coherent tunneling and keep the model minimal, we assume equal bare tunneling strengths to the two valley components, $\Gamma_{r{\rm K}}=\Gamma_{r{\rm K}'}\equiv\Gamma_r$. Thus, the difference between the tunneling rates of the two dot eigenstates originates solely from the coherent interference at the contact. Eqs.~\eqref{Gammaminus} and \eqref{Gammaplus} then reduce to
\begin{align}
\Gamma_{r\pm} 
=
\Gamma_r\left(1\mp C_r\sin\phi\right),
\label{eq:Gamma_effective}
\end{align}
where
\begin{align}
C_r=p_r\cos\Phi_r,
\qquad -1\le C_r\le1.
\end{align}
Equation~\eqref{eq:Gamma_effective} shows the essential consequence of valley-coherent tunneling. Although the bare tunneling strengths to the two valley components are identical, the valley-mixed eigenstates can have different tunneling rates, $\Gamma_{r+}\neq\Gamma_{r-}$.
This difference is largest when the valley mixing is strong, while it vanishes as $\sin\phi\rightarrow0$. As discussed below, this state-dependent tunneling provides a transport readout of the orbital-rotation-induced transitions between $|+\rangle$ and $|-\rangle$~\footnote{In general, the tunneling-rate matrix also has off-diagonal elements when expressed in the eigenstate basis, which can generate coherence between $|+\rangle$ and $|-\rangle$.
Here, however, we work in the secular regime in which the level splitting $\Delta E$ is much larger than the tunnel broadening. The off-diagonal elements of the reduced density matrix then rapidly oscillate and can be neglected in the population dynamics. For the representative parameters considered below, $\Delta E\simeq70\,\mu{\rm eV}$ whereas $\hbar(\Gamma_{\rm L}+\Gamma_{\rm R})\simeq2\,\mu{\rm eV}$, so that the secular approximation is well justified.}.

Within this approximation, the electron tunneling rates between lead \(r=L,R\) and the CNT quantum dot are calculated from $H_\text{lead--dot}$ as
\begin{align}
\Gamma^r_{(0,m)\rightarrow(\nu,m)} &= \Gamma_{r\nu} f_r (E_\nu), \label{eq:Gamma_tunnel_1} \\
\Gamma^r_{(\nu,m)\rightarrow(0,m)} &= \Gamma_{r\nu}\left[1-f_r (E_\nu)\right], \label{eq:Gamma_tunnel_2}
\end{align}
where $\nu=\pm$ labels the eigenstates of the CNT quantum dot, and $f_r(\epsilon)=[e^{\beta(\epsilon-\mu_r)}+1]^{-1}$ is the Fermi--Dirac distribution of electrode $r$, with $\beta=1/(k_{\mathrm B}T)$.
Ordinary tunneling does not change the phonon number $m$.

The damping of the torsional mode is described by
phonon relaxation and excitation rates, which are calculated from $H_{\text{ph--env}}$ as
\begin{align}
\Gamma^{\rm ph}_{(\alpha,m)\rightarrow(\alpha,m-1)} &= k_{\rm relax}\,m\,[1+n(\hbar \omega_0)], \\
\Gamma^{\rm ph}_{(\alpha,m-1)\rightarrow(\alpha,m)} &= k_{\rm relax}\,m\,n(\hbar \omega_0),
\end{align}
where \(n(\hbar \omega_0)\) is the Bose-Einstein distribution and \(k_{\mathrm{relax}}\) is the prefactor defined as 
\begin{align}
k_{\mathrm{relax}} = \frac{2\pi}{\hbar}\kappa^2 \mathcal{D}(\hbar \omega_0).
\end{align}
Here, \(\mathcal{D}(\hbar \omega_0) \) denotes the phonon density of states of the bath evaluated at the phonon energy \(\hbar\omega_0\).
At sufficiently low temperatures ($k_{\rm B} T \ll \hbar \omega_0$), $k_{\mathrm{relax}}$ is related to the quality factor of the torsional mode as $Q = \omega_0/k_{\mathrm{relax}}$.
Based on measurements of torsional and flexural modes~\cite{Hall2008, Huttel2009}, $Q$ is expected to range from $10^{2}$ to $10^{5}$ at low temperatures.

The orbital-rotation coupling in Eq.~\eqref{eq:H_OR_final} is diagonal in the valley basis.
Therefore, in the absence of intervalley mixing (\(\Delta_{{\rm KK}'} = 0\)), it only modulates the
energy splitting between \(K\) and \(K'\) and does not induce transitions between the energy eigenstates. 
For finite intervalley mixing, however, the situation changes when the magnetic field is tuned near the valley anticrossing~\footnote{
The intervalley mixing in the dot and the valley-coherent tunneling at the contacts therefore play distinct roles. The former introduces an off-diagonal component into the orbital-rotation interaction in the dot eigenbasis and thereby enables transitions between the two eigenstates, whereas the latter can make these eigenstates couple differently to the electrodes and convert the internal transitions into a change in the transport current.
}.

We introduce a unitary $2\times 2$ matrix \(\mathcal U_\phi \), which transforms the valley operator from the valley basis to the eigenbasis as follows:
\begin{align}
\mathcal U_\phi^\dagger \hat{\tau}_z \mathcal U_\phi
= \cos\phi \,\hat{\tau}'_z - \sin\phi\,\hat{\tau}'_x .
\end{align}
Here, \(\hat{\tau}'_{x,z}\) are Pauli matrices in the eigenstate basis \(\{|-\rangle,|+\rangle\}\). 
Applying this transformation, the orbital-rotation Hamiltonian in the eigenstate basis becomes
\begin{align}
H'_{\rm OR} &= \mathcal U_\phi^\dagger H_{\rm OR} \mathcal U_\phi \notag \\
&= ig_{\rm OR}(a-a^\dagger)
\left( \cos\phi \,\hat{\tau}'_z - \sin\phi \,\hat{\tau}'_x \right).
\label{eq:H_OR_eigenbasis}
\end{align}
Here, the second term induces transitions between \(|+\rangle\) and
\(|-\rangle\). 

The orbital-rotation-induced transitions are
\begin{align}
(+,m) &\rightarrow (-,m+1),
\\
(-,m) &\rightarrow (+,m-1),
\end{align}
corresponding respectively to phonon emission and absorption. 
These transition rates are calculated from $H_{\rm OR}$ as
\begin{align}
\Gamma^{\rm OR}_{(+,m)\rightarrow(-,m+1)} &= (m+1)\Gamma_{\rm OR}, \\
\Gamma^{\rm OR}_{(-,m)\rightarrow(+,m-1)} &= m\Gamma_{\rm OR},
\end{align}
where
\begin{align}
\Gamma_{\rm OR} &=  A\,\delta(\Delta E-\hbar\omega_0), \\
A &= \frac{8\hbar^2 l^2 \omega_0^2}{\pi^2 R^2} \sqrt{\frac{1}{GS\rho}} \sin^2\phi .
\label{eq:A_OR}
\end{align}
The mixing factor \(\sin^2\phi \) is maximal at the valley anticrossing (\(\delta B=0\)), whereas the full transition rate is also controlled by the resonance factor and is therefore largest near \(\Delta E=\hbar\omega_0\).
For numerical calculations, the delta function is replaced by a Lorentzian
with a lifetime broadening \(\hbar\Gamma\):
\begin{align}
\delta(\Delta E-\hbar\omega_0)
\rightarrow
\frac{\hbar\Gamma/2\pi}
{(\Delta E-\hbar\omega_0)^2+(\hbar\Gamma/2)^2}.
\end{align}
In the parameter regime considered below, the electronic lifetime broadening dominates over mechanical damping, and we take $\Gamma=\Gamma_{\rm L}+\Gamma_{\rm R}$.
To characterize the strength of the orbital-rotation coupling, we define the orbital-rotation transition rate at resonance as 
\begin{align}
\Gamma_{\rm OR}^{\rm res} \equiv \left. \Gamma_{\rm OR} \right|_{\Delta E=\hbar\omega_0} 
&= \frac{16 l^2 }{\pi^3\hbar\Gamma R^2 } \sqrt{\frac{1}{GS\rho}}\Delta_{{\rm KK}'}^2.
\label{eq:Gamma_OR_res}
\end{align}
We note that this rate is independent of the mixing angle $\phi$ (or the detuning field $\delta B$).

\subsection{Master equation}

The occupation probabilities \(P_{\alpha,m}\) obey the master equation
\begin{align}
\dot{P}_{0,m}
=&
\sum_{\substack{r = {\rm L}, {\rm R}\\\nu = +, -}}\left\{ - \Gamma^r_{(0,m)\rightarrow(\nu,m)}P_{0,m}+ \Gamma^r_{(\nu,m)\rightarrow(0,m)}P_{\nu,m}\right\}
\notag\\
&+{\cal D}[P_{0,m}],
\\[2mm]
\dot{P}_{+,m}
=&
-(m+1)\Gamma_{\rm OR} P_{+,m}+(m+1)\Gamma_{\rm OR} P_{-,m+1}
\notag\\
&+
\sum_{r = {\rm L},{\rm R}}\Bigl\{\Gamma^r_{(0,m)\rightarrow(+,m)}P_{0,m}\notag \\
& \hspace{10mm} -
\Gamma^r_{(+,m)\rightarrow(0,m)}P_{+,m}\Bigr\}
+{\cal D}[P_{+,m}],
\\[2mm]
\dot{P}_{-,m}
=&
-m\Gamma_{\rm OR} P_{-,m}+m\Gamma_{\rm OR} P_{+,m-1}
\notag\\
&+
\sum_{r = {\rm L},{\rm R}}\Bigl\{\Gamma^r_{(0,m)\rightarrow(-,m)}P_{0,m}
\notag\\
&\hspace{10mm} -
\Gamma^r_{(-,m)\rightarrow(0,m)}P_{-,m}\Bigr\}
+{\cal D}[P_{-,m}],
\end{align}
where
\begin{align}
{\cal D}[P_{\alpha,m}] 
&= k_{\mathrm{relax}}\Big[(m+1)(1+n)P_{\alpha,m+1}+m n P_{\alpha,m-1}
\notag\\
&\quad-m(1+n) P_{\alpha,m}-(m+1)n P_{\alpha,m}\Big],
\end{align}
and $n = n(\hbar \omega_0)$ is the Bose-Einstein distribution function.
The steady state is obtained from
$\dot P_{\alpha,m}=0$ 
in combination with the normalization condition 
$\sum_{\alpha=0,\pm}\sum_m P_{\alpha,m}=1$.
In numerical calculations, the phonon number is truncated at a sufficiently
large cutoff \(m_{\rm max}\).
The current flowing from the left electrode into the CNT quantum dot is calculated as
\begin{align}
I_L = e\sum_{m=0}^\infty \sum_{\nu=\pm}
\Bigl[
\Gamma^L_{(0,m)\rightarrow(\nu,m)}P_{0,m}  - \Gamma^L_{(\nu,m)\rightarrow(0,m)}P_{\nu,m}
\Bigr].
\end{align}
In the steady state, the current calculated at the right electrode gives the same value, up to the sign convention.

\section{Parameter estimate}
\label{sec:estimate}

We estimate the relevant experimental parameters for the proposed torsional drive.
We consider a single-wall CNT with a radius \( R \approx 0.68\,\mathrm{nm} \), a wall thickness \( d \approx 0.34\,\mathrm{nm} \), and a length \(L = 400\,\mathrm{nm} \).
The cross-sectional area \( S = 2\pi R d \approx 1.45 \times 10^{-18}\,\mathrm{m}^2 \) and the linear mass density \( \rho \approx 3.28 \times 10^{-15}\,\mathrm{kg}/\mathrm{m} \) are calculated from the mass density of graphene $2.26 \times 10^{3}\, \text{kg/m}^3$. 
We adopt a representative shear modulus of $G=0.4\,\mathrm{TPa}$, of the order expected for CNTs \cite{Lu1997, Hall2006, Hall2010Erratum}.
Substituting these values into the dispersion relation for torsional waves, we obtain the fundamental torsional frequency \(\omega_0\simeq1.1\times10^{11}\,{\rm rad/s}\), corresponding to the phonon energy \(\hbar\omega_0\simeq 70\,\mu{\rm eV}\).

The intervalley mixing \(\Delta_{{\rm KK}'}\) is a sample-dependent parameter because it originates from impurity scattering, disorder, and boundary effects in the CNT. Experimentally, Kuemmeth et al. reported a small disorder-induced intervalley mixing of \(\Delta_{{\rm KK}'}\simeq 65\,\mu{\rm eV}\) in an ultraclean CNT quantum dot~\cite{Kuemmeth2008}. 
The orbital magnetic moment is given by \(\mu_{\rm orb}\simeq e v_F R/2\)~\cite{Laird2015}, leading to $\mu_{\rm orb}\simeq0.27\,{\rm meV/T}$ for \(R=0.68\,{\rm nm}\).
Using these values, the valley splitting \(\Delta E \)
matches \(\hbar\omega_0\simeq70\,\mu{\rm eV}\) at a magnetic-field detuning \( |\delta B| \simeq 0.05\,{\rm T} \)
from the center of the valley anticrossing \(B_{\downarrow}\). 
We estimate the magnitude of the effective orbital angular momentum using the semiclassical relation $|\mu_{\rm orb}|\simeq e|L_z|/(2m_e)$.
With $|\mu_{\rm orb}|\simeq ev_FR/2$, this gives
\begin{equation}
|L_z|\simeq m_ev_FR.
\label{eq:Lz_estimate}
\end{equation}
For an order-of-magnitude estimate, we consider a gapped CNT with \(R\simeq0.68\,\mathrm{nm}\), in a regime where the active orbital lies sufficiently close to the band edge that the axial-confinement correction to the orbital magnetic moment is small. We then use \(\mu_{\rm orb}\simeq ev_FR/2\). 
This gives $|l|=|L_z|/\hbar\simeq4.7$ for $v_F=8\times10^5\,\mathrm{m/s}$.
This corresponds to an orbital angular momentum of approximately $4.7\hbar$, about nine times the electron spin angular momentum $\hbar/2$. We use the rounded value $|l|=5$ in the numerical estimates~\footnote{We stress that Eq.~\eqref{eq:Lz_estimate} relies on the free-electron gyromagnetic
relation with the bare electron mass $m_{e}$, which is not a controlled approximation
for a Bloch electron in a CNT. The value $|l|\simeq5$ should therefore be regarded as
an order-of-magnitude estimate, and, since $\Gamma_{\rm OR}^{\rm res}\propto l^{2}$
[Eq.~\eqref{eq:Gamma_OR_res}], the predicted rate is sensitive to it.
This uncertainty, however, does not obviously act against the proposed mechanism.
Orbital magnetic moments up to seven times as large as the semiclassical value
$ev_{F}R/2$ have been reported in CNT quantum dots and attributed to
electron--electron interactions~\cite{Island2018}. If the same gyromagnetic relation remains applicable to these interaction-enhanced moments, such an enhancement would imply a correspondingly larger $|l|$ and hence a larger $\Gamma_{\rm OR}^{\rm res}$. In this sense, the value adopted here
is a conservative choice.}.
By choosing \( \hbar\Gamma_{\mathrm{L}} = \hbar\Gamma_{\mathrm{R}} = 1.0\, \mu\mathrm{eV} \), the intervalley transition rate at the resonance condition $\Delta E = \hbar \omega_0$ is roughly estimated as \( \hbar\Gamma_{\rm OR}^{\rm res}\simeq 1.4 \times 10^{-1}\mu\mathrm{eV} \).
Within this semiclassical estimate of the rotational coupling, the proposed mechanism can drive the torsional mode at resonance ($\Delta E=\hbar\omega_0$), provided that mechanical damping is sufficiently weak.

\section{Numerical Results}
\label{sec:results}

We present numerical results for the current response and the nonequilibrium excitation of the CNT torsional mode induced by the orbital-rotation coupling.
Throughout this section, the shear modulus, the radius of the CNT, the intervalley mixing strength, and bare tunneling rates are taken as $G=0.4\, \text{TPa}$, $R=0.68 \, \mathrm{nm}$,  $\Delta_{{\rm KK}'}=65\, \mu\mathrm{eV}$, and \( \hbar\Gamma_{\mathrm{L}} = \hbar\Gamma_{\mathrm{R}} = 1.0\, \mu\mathrm{eV} \), respectively.

Unless otherwise stated, we set the contact-interference parameters to
$C_\mathrm{L}=-0.4$ and $C_\mathrm{R}=0.4$. According to Eq.~\eqref{eq:Gamma_effective}, this choice gives $\Gamma_{{\rm L}+}>\Gamma_{{\rm L}-}$ and $\Gamma_{{\rm R}-}>\Gamma_{{\rm R}+}$ for $\sin\phi>0$.
Thus, electron injection from the left electrode preferentially populates the $\ket{+}$ state, whereas electron extraction into the right electrode is faster from the $\ket{-}$ state.
The orbital-rotation-induced transition $\ket{+,m}\rightarrow\ket{-,m+1}$ therefore connects the preferential injection and extraction channels.
The opposite signs of $C_{\mathrm L}$ and $C_{\mathrm R}$ represent different coherent tunneling interference at the two contacts and do not require valley-polarized electrodes.

\subsection{Resonant current enhancement}
\label{sec:resonant_current}

\begin{figure}[t]
\centering
\includegraphics[width=8.5cm]{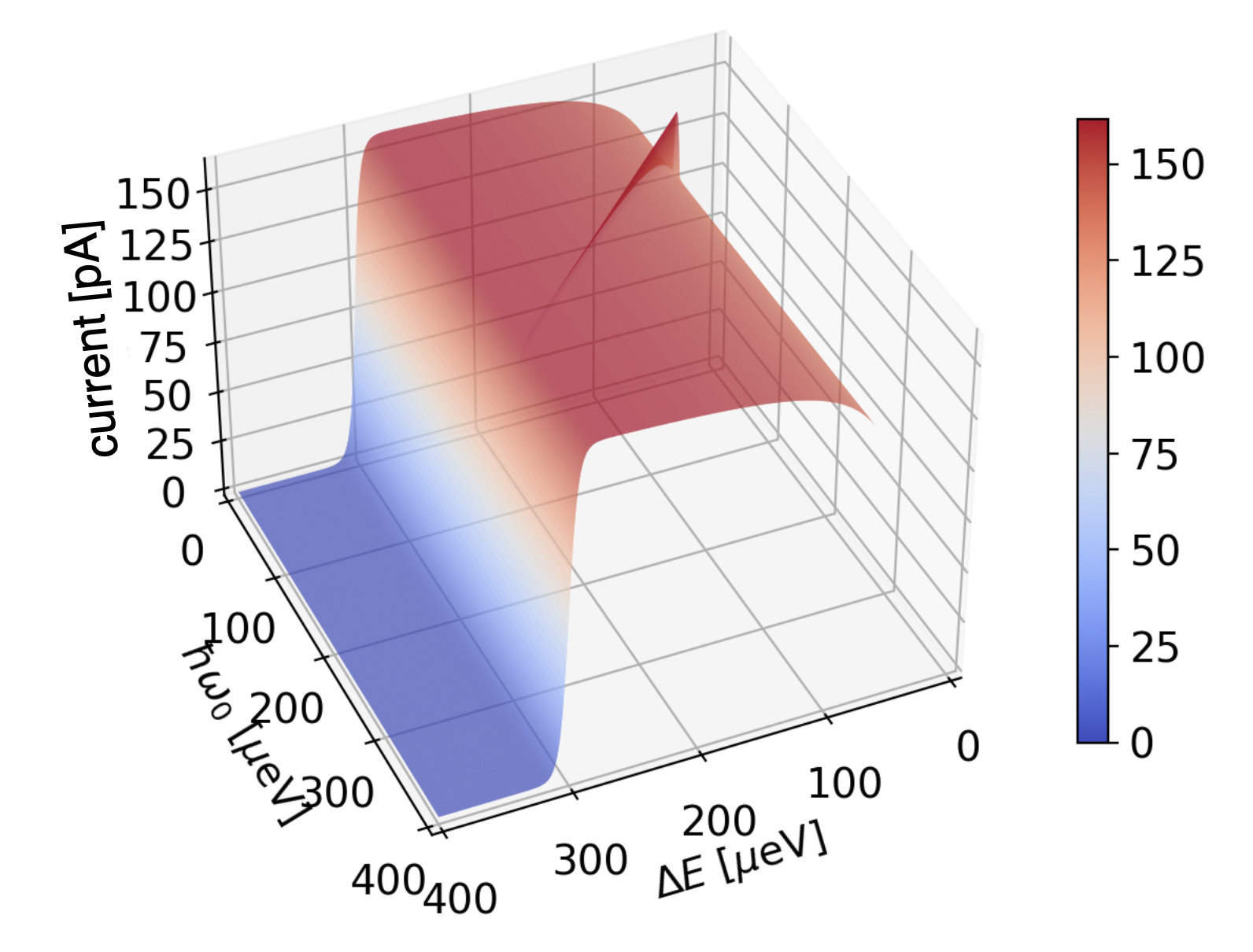}
\caption{Contour plot of the current as a function of $\hbar\omega_0$ and \(\Delta E\). 
The parameters are set to \( k_{\rm B}T = 2.0\,\mu\mathrm{eV}\) (\(T=23.2\,\mathrm{mK}\)), \( \mu_{\mathrm{L}} = -\mu_{\mathrm{R}} = 150\, \mu\mathrm{eV} \) and $k_\mathrm{relax} = 3.0\times10^{-5}\omega_0$.}
\label{fig:3d_current}
\end{figure}

Figure~\ref{fig:3d_current} shows the steady-state current as a function of the energy splitting \(\Delta E\) and the torsional phonon energy \(\hbar\omega_0\). For a fixed CNT radius \(R\), the phonon frequency \(\omega_0\) is controlled by changing the CNT length \(L\), whereas the energy splitting \(\Delta E\) can be tuned by varying the magnetic-field detuning \(\delta B\).
We set the chemical potentials as $\mu_{\mathrm{L}} = -\mu_{\mathrm{R}} = 150\,\mu$eV, for which the energy levels of the CNT, $E_+ = \Delta E/2$ and $E_- = -\Delta E/2$, are in the energy window $[\mu_{\mathrm{R}},\mu_{\mathrm{L}}]$, allowing electron transfer between the dot and the two electrodes when $\Delta E< 300\, \mu {\rm eV}$.
The other parameters are given in the caption of Fig.~\ref{fig:3d_current}.
A sharp ridge appears along the diagonal line \( \Delta E = \hbar\omega_0 \), where the energy difference between the \(\ket{+}\) and \(\ket{-}\) states matches the phonon energy. 
Under this resonant condition, the orbital-rotation coupling efficiently drives the phonon-emission process
$\ket{+,m}\rightarrow\ket{-,m+1}$.
For $C_{\mathrm R}=-C_{\mathrm L}=0.4$, this transition transfers an electron from the state preferentially populated by the left electrode to the state that couples more strongly to the right electrode. The resulting transport cycle,
\begin{equation}
\ket{0,m}
\rightarrow
\ket{+,m}
\rightarrow
\ket{-,m+1}
\rightarrow
\ket{0,m+1},
\end{equation}
relieves the contact-induced transport bottleneck and produces the resonant enhancement of the current.
The background current itself depends smoothly on the energy splitting $\Delta E = 2E_+$, because the state-dependent tunneling rates contain $\sin\phi=\Delta_{{\rm KK}'}/(2 E_+)$.
For the present choice $C_{\mathrm R}=-C_{\mathrm L}$, the mismatch between the preferred injection and extraction channels is strongest near the intervalley anticrossing and decreases with increasing $\Delta E$. For example, at $\Delta E=\hbar\omega_0=70\,\mu\mathrm{eV}$, the current increases from approximately $134\,\mathrm{pA}$ in the absence of the orbital-rotation coupling to approximately $160\,\mathrm{pA}$ at resonance.

The resonant peak current reaches a value close to the maximum background current in the high-bias regime, where both dot levels lie well inside the bias window and $f_{\mathrm L}(E_\pm)\simeq1$ and $f_{\mathrm R}(E_\pm)\simeq0$.
In this limit, the total electron-injection rate from the left electrode is $\Gamma_{{\rm L}+}+\Gamma_{{\rm L}-}=2\Gamma_{\rm L}$ from Eq.~\eqref{eq:Gamma_effective}.
At resonance, the orbital-rotation-induced transitions redistribute the occupation between the $\ket{+}$ and $\ket{-}$ states and reduce the contact-induced population imbalance.
We denote the electronic occupation probabilities by $P_\alpha=\sum_m P_{\alpha,m}$, with $\alpha=0,+,-$.
When the two occupations become approximately equal, $P_+\simeq P_-$, the total extraction rate into the right electrode becomes approximately $\Gamma_{\rm R}(P_++P_-)=\Gamma_{\rm R}(1-P_0)$, since $\Gamma_{{\rm R}+}+\Gamma_{{\rm R}-}=2\Gamma_{\rm R}$.
The stationary condition therefore gives
\begin{equation}
2\Gamma_{\rm L}P_0
\simeq
\Gamma_{\rm R}(1-P_0),
\end{equation}
and hence the resonant current approaches
\begin{equation}
I_{\rm res}
\simeq
\frac{2e\Gamma_{\rm L}\Gamma_{\rm R}}
{2\Gamma_{\rm L}+\Gamma_{\rm R}}.
\end{equation}
For the symmetric coupling used here, $\Gamma_{\rm L}=\Gamma_{\rm R}=\Gamma_0$, this gives $I_{\rm res}\simeq2e\Gamma_0/3$~\footnote{The background current approaches the same value away from the intervalley anticrossing, as long as both dot levels remain well inside the bias window. With increasing $\Delta E=2E_+$, $\sin\phi=\Delta_{{\rm KK}'}/(2E_+)$ decreases, so that the difference between $\Gamma_{r+}$ and $\Gamma_{r-}$ becomes small and the contact-induced transport bottleneck disappears even without the orbital-rotation coupling. Thus, the near coincidence between the resonant peak current and the maximum background current has a simple interpretation: in both cases, the state-dependent tunneling bottleneck is effectively removed, either dynamically by the orbital-rotation-induced transitions at resonance or by the suppression of the contact-induced state selectivity away from the anticrossing.}.

\subsection{Phonon excitation}
\label{sec:torsional_excitation}

\begin{figure}[tb]
\centering
\includegraphics[width=8.5cm]{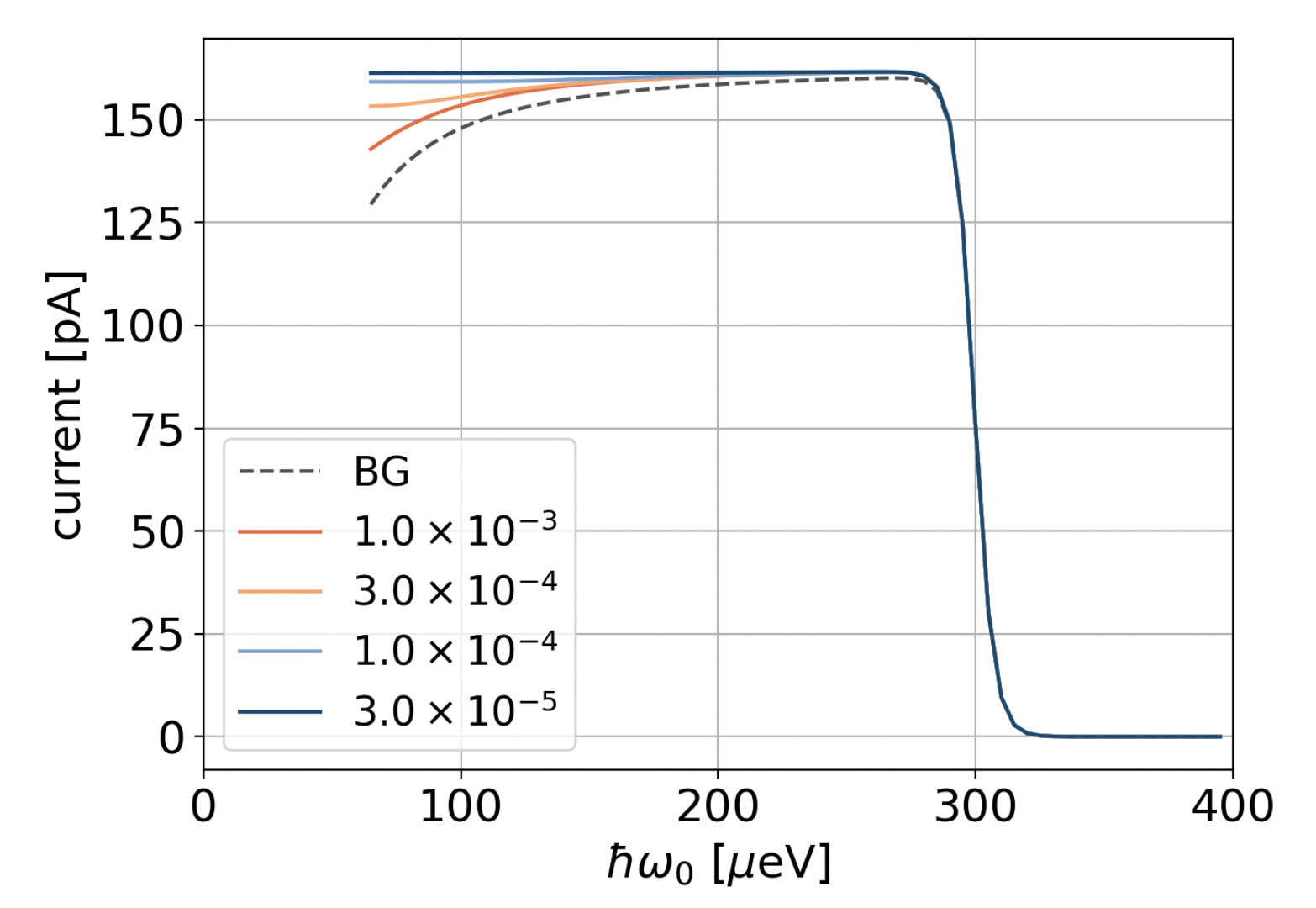}
\caption{The current under the resonant condition ($\Delta E=\hbar \omega_0$) is plotted as a function of $\hbar\omega_0$ at \( k_{\mathrm{B}} T = 2.0\,\mu\mathrm{eV}\) (\(T=23.2\,\mathrm{mK}\)). The chemical potentials are $\mu_{\mathrm L}=-\mu_{\mathrm R}=150\,\mu\mathrm{eV}$.
The four solid curves correspond to $k_{\rm relax}/\omega_0 = 3\times10^{-5}$, $1\times10^{-4}$, $3\times10^{-4}$, and $1\times10^{-3}$, respectively.
The dashed curve represents the result without the orbital-rotation coupling ($\Gamma_{\rm OR}=0$).}
\label{fig:2d_current_krelax_T02_10}
\end{figure}

Figure~\ref{fig:2d_current_krelax_T02_10} shows the current along the resonance line $\Delta E=\hbar\omega_0$ for several values of the ratio $k_{\rm relax}/\omega_0$, which corresponds to the inverse quality factor of the torsional mode.
The dashed line shows the current without the orbital-phonon coupling.
We find that the current enhancement due to the orbital-rotation coupling, i.e., the height of the ridge in Fig.~\ref{fig:3d_current}, is reduced by increasing $k_{\rm relax}/\omega_0$ and the current approaches the background current denoted by the dashed line since the phonon damping suppresses the nonequilibrium phonon population.

\begin{figure}[tb]
\centering
\includegraphics[width=8.5cm]{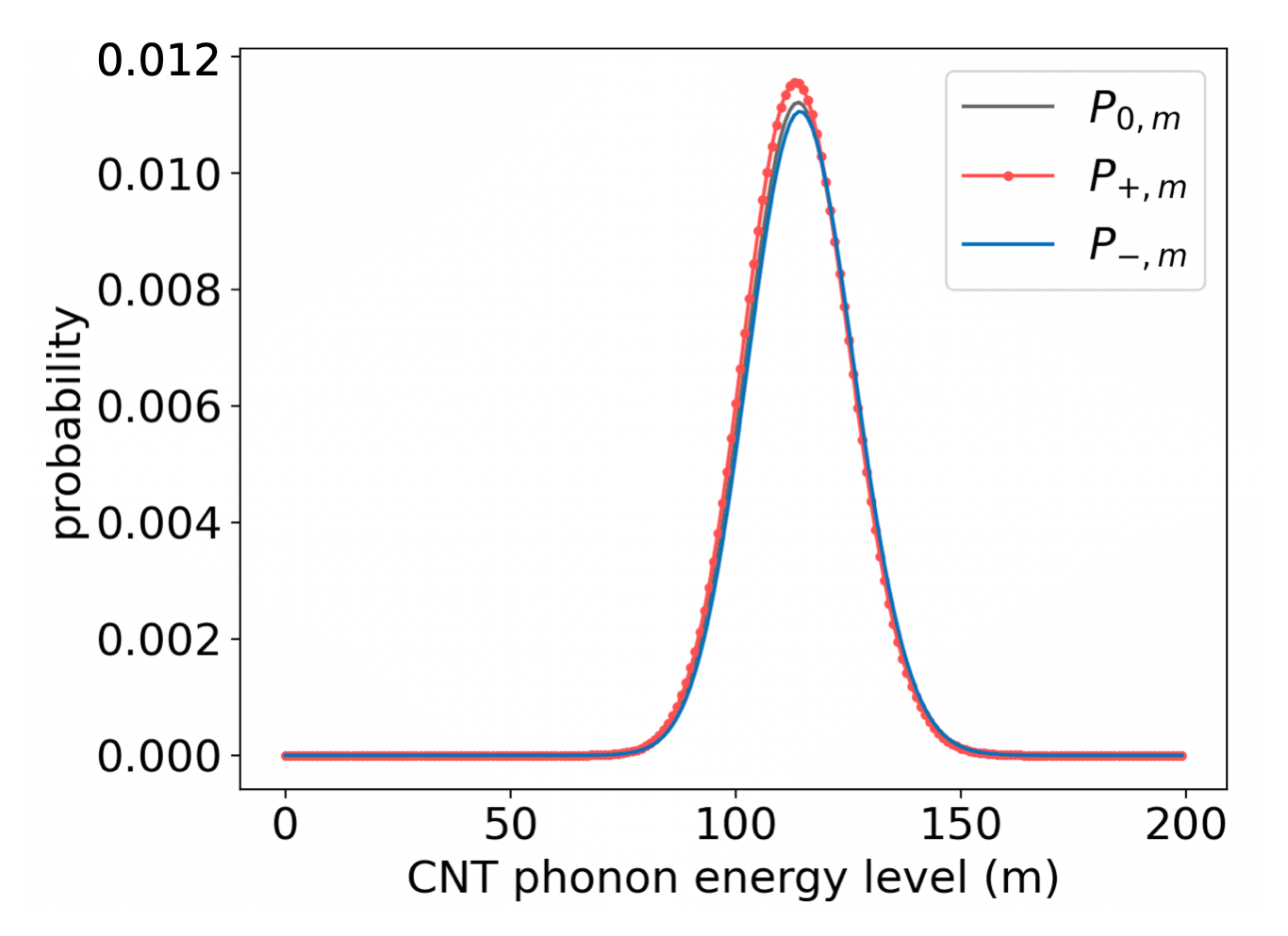}
\caption{Probability distributions of the CNT torsional mode. The parameters are set to \( \Delta E=\hbar\omega_0 = 70\, \mu\mathrm{eV} \), \( k_{\mathrm{B}}T = 2.0\,\mu\mathrm{eV}\) (\(T=23.2\,\mathrm{mK}\)), \( \mu_{\mathrm{L}} = -\mu_{\mathrm{R}} = 150\, \mu\mathrm{eV} \) and \( k_\mathrm{relax} = 3.0\times10^{-5}\omega_0 \).}
\label{fig:2d_phonon_dist}
\end{figure}

Figure~\ref{fig:2d_phonon_dist} shows the phonon number distribution of the CNT torsional mode at resonance \( \Delta E=\hbar\omega_0 = 70\, \mu\mathrm{eV} \), where the other parameters are set to \( k_{\mathrm{B}}T = 2.0\,\mu\mathrm{eV}\) (\(T=23.2\,\mathrm{mK}\)), \( \mu_{\mathrm{L}} = -\mu_{\mathrm{R}} = 150\, \mu\mathrm{eV} \), \( k_\mathrm{relax} = 3.0\times10^{-5}\omega_0 \) and $C_\mathrm{R} = -C_\mathrm{L} = 0.4$.
The distribution is strongly displaced from its thermal equilibrium form and has a maximum around $m\simeq115$, demonstrating substantial nonequilibrium excitation of the torsional mode.
Since $n(\hbar\omega_0)\ll1$ at this temperature, this large phonon occupation cannot be attributed to thermal excitation.
We note that the resulting average phonon number is larger than that obtained in the spin-rotation-based driving mechanism proposed in Ref.~\cite{Yamada2026}, for the respective parameter sets used.
The peak position can be estimated from the balance between orbital-rotation-induced pumping and mechanical relaxation using the master equation. In the high-bias limit with \(\Gamma_{\rm L}=\Gamma_{\rm R}=\Gamma_0\) and \(C_{\rm R}=-C_{\rm L}=C\), the net phonon pumping rate approaches \(J_{\rm pump} \simeq \Gamma_{{\rm L}+}P_0 - \Gamma_{{\rm R}+}P_+ \) when \((m+1)\Gamma_{\rm OR}\gg\Gamma_0\).
Using $P_+ \simeq P_- \simeq P_0 \simeq 1/3$ and balancing this rate with the mechanical relaxation rate \(k_{\rm relax}m\), we obtain
\begin{align}
m_{\rm peak}\simeq \frac{2\Gamma C\sin\phi}{3k_{\rm relax}} \simeq118, 
\end{align}
which agrees well with the numerical maximum around \(m\simeq115\).

\subsection{Contact dependence}
\label{sec:contact_dependence}

\begin{figure}[tb]
\centering
\includegraphics[width=8.5cm]{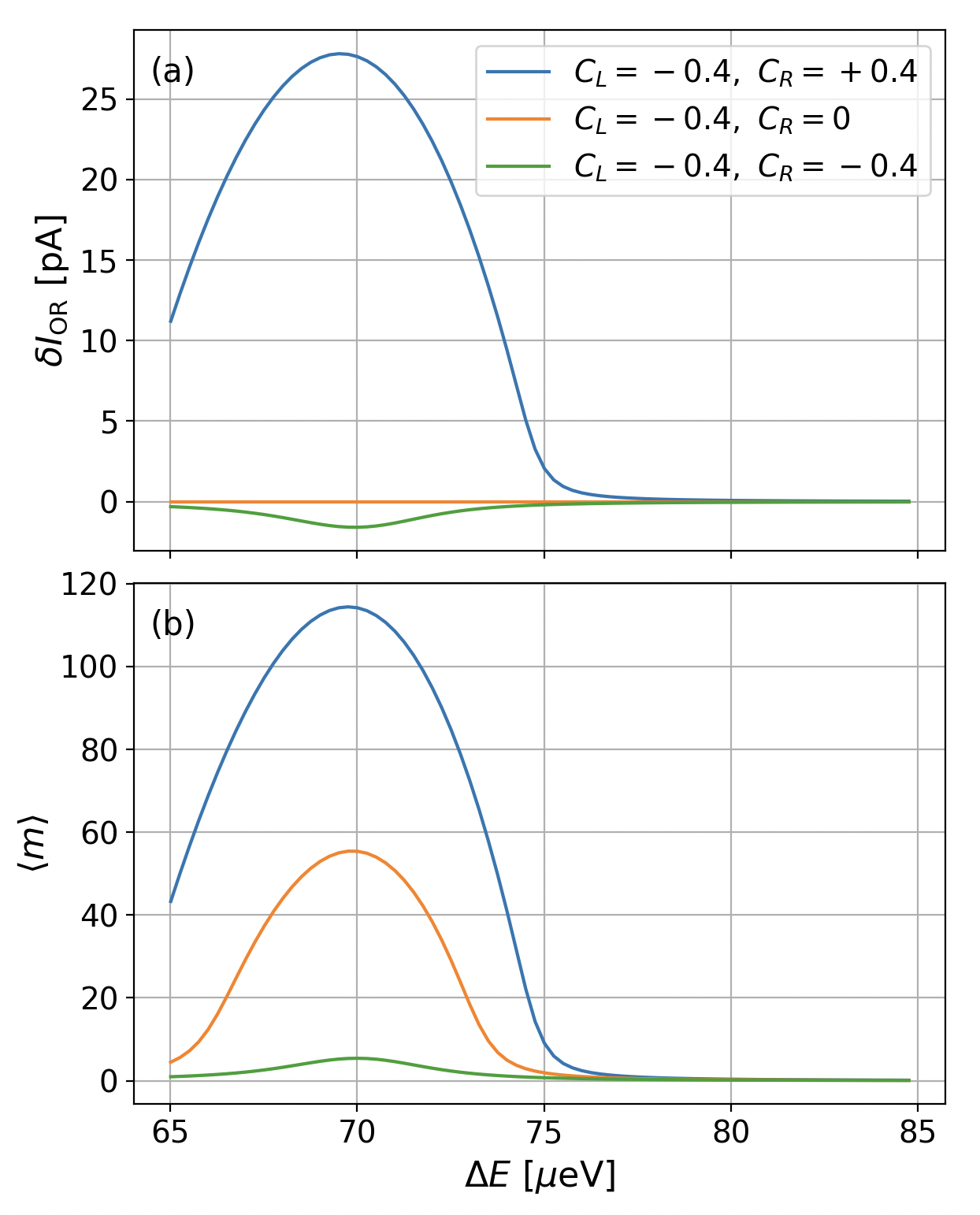}
\caption{(a) Current change due to the resonant phonon population, defined as $\delta I_{\rm OR}(\Delta E) = I(\Delta E)-I_0(\Delta E)$, where $I_0(\Delta E)$ is the background current in the absence of the orbital-rotation coupling ($\Gamma_{\rm OR}=0$).
The three curves correspond to $(C_{\mathrm L},C_{\mathrm R})=(-0.4,0.4)$, $(-0.4,0)$, and $(-0.4,-0.4)$, respectively.
(b) Corresponding average phonon number $\langle m\rangle$ of the CNT torsional mode. 
The parameters are taken as $\hbar\omega_0=70\,\mu\mathrm{eV}$,
$k_{\rm B}T=2.0\,\mu\mathrm{eV}$ ($T=23.2\,\mathrm{mK}$), $\mu_{\mathrm L}=-\mu_{\mathrm R}=150\,\mu\mathrm{eV}$, $k_{\rm relax}/\omega_0=3.0\times10^{-5}$, and $\hbar\Gamma_{\mathrm L}=\hbar\Gamma_{\mathrm R}=1.0\,\mu\mathrm{eV}$.}
\label{fig:contact_dependence}
\end{figure}

Finally, we discuss how the results depend on the contact parameters $C_{\rm L}$ and $C_{\rm R}$.
Figure~\ref{fig:contact_dependence}(a) shows the current change, defined as $\delta I_{\rm OR}(\Delta E) = I(\Delta E)-I_0(\Delta E)$, for $(C_{\rm L},C_{\rm R}) = (-0.4,0.4)$, $(-0.4,0)$, and $(-0.4,-0.4)$, where $I_0(\Delta E)$ is the background current in the absence of orbital-rotation coupling ($\Gamma_{\rm OR} = 0$).
A positive resonant signal appears for $(C_{\rm L},C_{\rm R}) = (-0.4,0.4)$, whereas a weaker negative signal appears for $(C_{\rm L},C_{\rm R}) = (-0.4,-0.4)$.
For $C_{\rm R}=0$, no resonant change in the current is observed.
These behaviors can be understood as follows.
For $C_{\rm L}<0<C_{\rm R}$, the tunneling rates from the left and right leads satisfy $\Gamma_{{\rm L}+} > \Gamma_{{\rm L}-}$ and $\Gamma_{{\rm R}+} < \Gamma_{{\rm R}-}$.
In this case, the phonon-assisted transition from $\ket{+}$ to $\ket{-}$ increases the current because it allows electrons to bypass the transport paths with small tunneling rates, $\Gamma_{{\rm L}-}$ and $\Gamma_{{\rm R}+}$.
In contrast, when $C_{\rm L},C_{\rm R}<0$, the same transition decreases the current.
Note that for $C_{\rm R}=0$, the current remains unchanged even though the torsional mode is still excited (see the discussion below).

Figure~\ref{fig:contact_dependence}(b) shows the average phonon number $\langle m \rangle$ as a function of $\Delta E$ for $(C_{\rm L},C_{\rm R}) = (-0.4,0.4)$, $(-0.4,0)$, and $(-0.4,-0.4)$.
For all the cases, the population reaches a maximum at the resonance condition $\Delta E=\hbar\omega_0=70\,\mu\mathrm{eV}$.
The resonance widths are of order $5\,\mu\mathrm{eV}$ and depend on the transition rates $\Gamma_{\rm L}$ and $\Gamma_{\rm R}$ between the electrodes and the CNT quantum dot, as well as on the contact configuration.
The average phonon number is largest for $(C_{\rm L},C_{\rm R}) = (-0.4,0.4)$ and smallest for $(C_{\rm L},C_{\rm R}) = (-0.4,-0.4)$.
Note that the phonon number remains finite for $(C_{\rm L},C_{\rm R}) = (-0.4,0)$.
These results can be explained by the imbalance in the occupation probabilities of the two eigenstates in the CNT quantum dot.
In the absence of orbital-rotation coupling ($\Gamma_{\rm OR} = 0$), the higher-energy state $\ket{+}$ has a larger (smaller) occupation probability than the lower-energy state $\ket{-}$ for $C_{\rm R}>C_{\rm L}$ ($C_{\rm R}<C_{\rm L}$).
Since the rate of the phonon-assisted transition from $\ket{+}$ to $\ket{-}$ is proportional to the occupation probability of $\ket{+}$, the phonon number is enhanced for $C_{\rm R}>C_{\rm L}$.

The parameters $C_r$ depend on the contact geometry and the CNT electronic states. Further microscopic studies of their relation to CNT boundary conditions and contact geometry would help clarify how favorable contact configurations can be realized.

\section{Relation to the gyromagnetic effects}
\label{sec:discussion}

Our mechanism is related to gyromagnetic angular-momentum conversion, but the angular momentum is carried by valley orbital motion rather than by spin.
Torque generation by gyromagnetic effects has been discussed since the classical works on the Barnett and Einstein--de Haas effects~\cite{Barnett1915,Einstein1915,Barnett1935,Matsuo2011}.
The conversion between spin angular momentum and mechanical rotation has attracted renewed attention in the 21st century in the context of spintronics~\cite{Matsuo2013,Kobayashi2017,Kurimune2020,Funato2025}.
Related studies include nanomechanical detection of spin angular momentum~\cite{Mohanty2004,Zolfagharkhani2008}, mechanical responses associated with spin-current injection~\cite{Harii2019}, and phonon angular momentum induced by a temperature gradient~\cite{Hamada2018}.
In a CNT quantum dot, the spin-phonon conversion has also been utilized for the Einstein--de Haas nanorotor~\cite{Izumida2022} and torsional mode excitation~\cite{Yamada2026}.

In these applications, spin relaxation transfers spin angular momentum to mechanical angular momentum.
In the present mechanism, the corresponding internal degree of freedom is the valley orbital motion in a CNT.
The intervalley mixing caused by disorder or tube-end scattering is essential because it converts the orbital-rotation coupling, which is diagonal in the valley basis, into an off-diagonal coupling between the valley-mixed eigenstates.
Equivalently, the mechanical rotation acts as a time-dependent detuning field conjugate to the valley orbital angular momentum.
The same viewpoint gives a route to angular-momentum conversion in solids beyond spin-based mechanisms.
Our results suggest that the conversion of orbital angular momentum into mechanical rotation, analogous to the mechanism originally envisioned by Einstein and de Haas, deserves renewed attention.

\section{Summary}
\label{sec:summary}

In summary, we have investigated a mechanism for driving a torsional mode of a suspended CNT via the orbital angular momentum of electrons in a CNT quantum dot. The key ingredient is the orbital-rotation coupling between the angular velocity of the torsional mode and the valley orbital angular momentum. In the presence of intervalley mixing, the dot eigenstates are superpositions of the two valley states, and the orbital-rotation coupling induces transitions between these eigenstates with emission or absorption of torsional phonons.

We formulated the coupled dynamics of electron transport and the torsional mode using a master equation in the sequential-tunneling regime. Although the electrodes are normal metals without intrinsic valley polarization, coherent tunneling at the contacts can couple differently to the two valley-mixed eigenstates of the CNT. This state-dependent tunneling provides a transport readout of the orbital-rotation-induced transitions. 
At resonance, when the energy splitting \( \Delta E = E_{+}-E_{-}\) between the valley-mixed states is tuned close to the torsional phonon energy \( \hbar\omega_0 \), phonon-assisted orbital transitions become resonant and modify the steady-state transport. 
The resulting current response depends on the contact configuration and can be positive, vanishing, or negative, even when the torsional mode is excited.

The proposed mechanism provides a route to actuate CNT torsional motion through orbital angular-momentum transfer without the ferromagnetic electrodes required in the spin-based setup~\cite{Yamada2026}. Because the orbital angular momentum associated with CNT valley states can be much larger than the spin angular momentum, the resulting coupling can be stronger than that based on the spin-rotation mechanism. These results suggest that CNT quantum dots are useful for converting electronic orbital angular momentum into mechanical angular momentum in nanoscale torsional resonators.

\begin{acknowledgments}
This work was supported by the Priority Program of Chinese Academy of Sciences under Grant No. XDB28000000, and by JSPS KAKENHI Grant Nos.~JP24K06951, JP21K03414 from MEXT, Japan.
\end{acknowledgments}

\bibliography{ref}

\end{document}